\documentclass[preprintnumbers,10pt,nofootinbib]{revtex4}

\usepackage{amsmath,latexsym,amssymb,amsfonts}
\usepackage[dvips]{graphicx,color}
\usepackage{bm}
\usepackage{cancel}

\begin{document}

\preprint{YITP-15-109}

\title{\textbf{Stationary bubbles and their tunneling channels toward trivial geometry}}

\author{
\textsc{Pisin Chen$^{a,b,c,d}$}\footnote{{\tt pisinchen{}@{}phys.ntu.edu.tw }},
\textsc{Guillem Dom\`enech$^{e}$}\footnote{{\tt  guillem.domenech{}@{}yukawa.kyoto-u.ac.jp}},
\textsc{Misao Sasaki$^{e}$}\footnote{{\tt misao{}@{}yukawa.kyoto-u.ac.jp}} 
and
\textsc{Dong-han Yeom$^{a}$}\footnote{{\tt innocent.yeom{}@{}gmail.com}}
}

\affiliation{
$^{a}$\small{Leung Center for Cosmology and Particle Astrophysics, National Taiwan University, Taipei 10617, Taiwan}\\
$^{b}$\small{Department of Physics, National Taiwan University, Taipei 10617, Taiwan}\\
$^{c}$\small{Graduate Institute of Astrophysics, National Taiwan University, Taipei 10617, Taiwan}\\
$^{d}$\small{Kavli Institute for Particle Astrophysics and Cosmology, SLAC National Accelerator Laboratory, Stanford University, Stanford, California 94305, USA}\\
$^{e}$\small{Yukawa Institute for Theoretical Physics, Kyoto University, Kyoto 606-8502, Japan}
}

\begin{abstract}
In the path integral approach, one has to sum over all histories that start from the same initial condition in order to obtain the final condition as a superposition of histories. Applying this into black hole dynamics, we consider stable and unstable stationary bubbles as a reasonable and regular initial condition. We find examples where the bubble can either form a black hole or tunnel toward a trivial geometry, i.e., with no singularity nor event horizon. We investigate the dynamics and tunneling channels of true vacuum bubbles for various tensions. In particular, in line with the idea of superposition of geometries, we build a classically stable stationary thin-shell solution in a Minkowski background where its fate is probabilistically given by non-perturbative effects. Since there exists a tunneling channel toward a trivial geometry in the entire path integral, the entire information is encoded in the wave function. This demonstrates that the unitarity is preserved and there is no loss of information when viewed from the entire wave function of the universe, whereas a semi-classical observer, who can see only a definitive geometry, would find an effective loss of information. This may provide a resolution to the information loss dilemma. 
\end{abstract}

\maketitle

\newpage

\tableofcontents


\section{Introduction}

Recently, there have been interesting discussions on the information loss problem of black holes \cite{Hawking:1976ra}. It has been clearly exposed that one cannot consistently hold on to general relativity, semi-classical quantum field theory, area-entropy relation, 
and unitarity all at the same time \cite{Yeom:2008qw,Almheiri:2012rt}. This means that for a certain stage some of the well-established physical principles should breakdown, not only at singularities but also at very large scales. Among various ideas to address this problem, there is a proposal to drop general relativity introducing the so-called firewall \cite{Almheiri:2012rt}. However, if there is a firewall, then it should affect the future infinity and hence that seems very problematic \cite{Hwang:2012nn,CNOPSY}.

An interesting approach to investigate this problem is via the wave function of the universe that satisfies the Wheeler-DeWitt equation \cite{Hartle:1983ai}. If the universe is in the ground state, then the wave function can be computed by an Euclidean path-integral. 
One can approximate the path-integral as a sum over on-shell histories
which are either Lorentzian or Euclidean, where the latter are called instantons.
Namely an instanton is a classical trajectory over the complexified time that 
connects the initial state to one of the final states.
 For a given initial state, the wave function behaves as a propagator and the final 
state will be a superposition of various semi-classical states. 

According to Maldacena \cite{Maldacena:2001kr} and Hawking \cite{Hawking:2005kf}, if one of the histories has a trivial topology, namely that there is no horizon nor singularity in the Lorentzian sense, then the information will be eventually conserved through the history, even though the probability is suppressed on the order of the entropy. This might explain why there is no information loss for the entire wave function of the universe, yet a semi-classical observer, who can see a definite geometry, would effectively detect the loss of information. It might well be that a unitary observer, if any, who lives in the superspace, could approximately describe classical observables as expectation values, where these expectation values do not have to satisfy the classical equations of motion. In this sense, there could be a violation of general relativity. 
But in this scenario there is no physical firewall for a semi-classical 
observer \cite{Sasaki:2014spa}.

While this framework from the path-integral, so-called the \textit{effective loss of information}, is very appealing, more justifications are needed. Possible issues of the effective loss of information are the following:
\begin{itemize}
\item[1.] Is the existence of an instanton that mediates a trivial geometry \textit{generic}?
\item[2.] Can one technically demonstrate the \textit{expectation values} of observables in terms of a unitary observer?
\item[3.] Is the existence of a trivial geometry really \textit{sufficient} to explain unitarity?
\end{itemize}
If we can answer these questions positively, the effective loss of information 
will be a genuine answer for the information loss paradox.

In this paper, we want to address the first question: \textit{is the existence of such an instanton generic}? Up to now, we are not yet able to answer this question, 
but we can at least find interesting and quite generic examples regarding 
this within the thin-shell approximation \cite{Sasaki:2014spa,Lee:2015rwa}. 
If a shell is self-gravitating and has a time-symmetric solution, then there is a hope to find a tunneling channel that links a collapsing shell to a bouncing shell toward a trivial geometry \cite{Farhi:1989yr}. In order to obtain a trivial geometry within the thin-shell approximation, we need the following conditions:
\begin{itemize}
\item[-- C1.] There should be time-symmetric collapsing and bouncing solutions.
\item[-- C2.] At the bouncing point, the extrinsic curvature of the shell should be positive.
\item[-- C3.] The shell should start with a regular initial condition and approach to a time-symmetric collapsing solution.
\end{itemize}
The existence of thin-shell solutions that satisfy C1 and C2 has already been shown in the literature \cite{Sasaki:2014spa,Gregory:2013hja}, while C3 was simply put in by hand. However, in order to make our argument more generic, we need to weaken these conditions. Regarding C3, for realistic star interiors, there exist various repulsive effects that can be well approximated by a modification of the tension term (for more detailed analytic descriptions of gravitational collapses, see \cite{Adler:2005vn}). In this paper, by modifying the tension around the star regime, we see that the realization of a regular initial condition that smoothly approaches a maximum radius is definitely reasonable and possible. 

This paper is organized as follows. In SEC.~\ref{sec:bri}, we briefly summarize the idea on the effective loss of information. In SEC.~\ref{sec:dyn}, we investigate true 
vacuum bubbles in asymptotic anti-de Sitter (AdS) and Minkowski backgrounds.
We construct some examples that have stable or unstable stationary points 
by varying the tension term. 
Finally, in SEC.~\ref{sec:con}, we summarize our results and discuss on future 
topics. 
Some useful but supplementary analyses are given in Appendix.
Throughout this paper, we assume the units $c = \hbar = G = 1$,
 while in some special cases we explicitly recover the constants.

\section{\label{sec:bri}Brief review on the effective loss of information}

The propagator from the initial state $|i\rangle$ to the final state $|f\rangle$ can be described by the path integral
\begin{eqnarray}
\langle f | i \rangle = \int \mathcal{D}g_{\mu\nu} \mathcal{D}\phi \; e^{iS[g_{\mu\nu},\phi]},
\end{eqnarray}
where $g_{\mu\nu}$ is the metric, $\phi$ is a matter field, and $S$ is the Lorentzian action. The final state, which is located at future infinity, is represented by a superposition of classical states $|f_{j}\rangle$, that is $|f\rangle = \sum_{j} a_{j} |f_{j}\rangle$, if one assumes the classicality of the wave function for an asymptotic observer.

In general, this Lorentzian path integral badly diverges. However, according to Hartle and Hawking \cite{Hartle:1983ai}, an analytic continuation of the time from Lorentzian signatures to Euclidean signatures will improve its behavior and the analytically continued path integral will correspond to the ground state wave function. This Euclidean path integral is presented by
\begin{eqnarray}
\langle f | i \rangle = \int \mathcal{D}g_{\mu\nu} \mathcal{D}\phi \; e^{-S_{\mathrm{E}}[g_{\mu\nu},\phi]},
\end{eqnarray}
where now $S_{\mathrm{E}}$ is the Euclidean action and we sum over all Euclidean paths that connect $|i\rangle$ to $|f_{j}\rangle$. Still, the whole path integral is difficult to solve exactly. Nevertheless, one can extract information about the ground state of the wave function by using the steepest-descent approximation, since in a semiclassical regime, Euclidean classical solutions (instantons) give the dominant contribution. In this approximation, one sums over instantons, namely
\begin{eqnarray}
\langle f | i \rangle \simeq \sum_{i \rightarrow f_{j}} e^{-S^{\mathrm{on-shell}}_{\mathrm{E}}}.
\end{eqnarray}
Thus, the exponential gives a rough estimation on the tunneling rate from $|i\rangle$ to $|f_{j}\rangle$.

To be concrete, imagine a situation where the initial state defined on the past infinity $|i\rangle$ contains a collapsing matter that will classically form a black hole, and let us call it $|f_{\mathrm{BH}}\rangle$. Note that under the path integral approach, $|f_{\mathrm{BH}}\rangle$ will not be in general the unique final state; in other words, $|i\rangle$ may undergo quantum tunneling. Ideally, if there exists a history with a trivial geometry, say from $|i\rangle$ to $|f_{\cancel{\mathrm{BH}}}\rangle$, which has no singularity nor event horizon, then every information of the initial state $|i\rangle$ will be conserved through this history. In fact, this point was previously discussed by Maldacena \cite{Maldacena:2001kr} and Hawking \cite{Hawking:2005kf}. The main idea is as follows: although the probability for $|f_{\cancel{\mathrm{BH}}}\rangle$ is exponentially suppressed compared to $|f_{\mathrm{BH}}\rangle$, the two point correlation function between inside and outside the horizon does not decay to zero. Thus, information will be eventually recovered through the trivial geometry \cite{Maldacena:2001kr}.

If this idea is general enough, then we might say that a unitary observer who can see the superposition of states $|f\rangle$ will restore all information, while a semiclassical observer will see a particular classical state, say $|f_{j}\rangle$, and will effectively lose information. This is the reason why this approach is called the \textit{effective loss of information}. It should be noted that a unitary observer need not satisfy classical equations of motion (general relativity included). 

Now, the question is whether the previous idea could be applicable for generic matter-gravity systems. We devote the following sections to find concrete examples where an instanton connects a collapsing matter to a trivial geometry.

\section{\label{sec:dyn}Dynamics and tunneling of true vacuum bubbles}

We consider a spherical thin-shell located at $r$ that divides two spacetime regions with the spherical symmetric metric ansatz (for a review of the thin-shell formalism, see \cite{Toolkit})
\begin{eqnarray}
\label{eq:metric}
ds_{\pm}^{2}= - f_{\pm}(R) dT^{2} + \frac{1}{f_{\pm}(R)} dR^{2} + R^{2} d\Omega^{2}\,,
\end{eqnarray}
where $\pm$ refer to outside ($R>r$) and inside ($R<r$) the shell, respectively.
The intrinsic metric of the thin-shell is given by
\begin{eqnarray}
ds^{2} = - dt^{2} + r^{2}(t) d\Omega^{2}.
\end{eqnarray}
We take the metric ansatz for outside and inside the shell to be
\begin{eqnarray}
f_{\pm}(R) = 1 - \frac{2M_{\pm}}{R} + \frac{R^{2}}{\ell_{\pm}^{2}},
\end{eqnarray}
where $M_{+}$ and $M_{-}$ are the mass parameters of each region and
\begin{eqnarray}
\ell^{2}_{\pm} = \frac{3}{8\pi \left| V_{\pm} \right|}
\end{eqnarray}
is the parameter due to the vacuum energy $V_{\pm}$. In this section, 
we focus on Minkowski and AdS background and consider true vacuum bubbles, but it can be easily generalized to a dS background. For our purposes we assume $M_{-}= 0$, unless stated otherwise, and therefore the internal manifold does not have a singularity.

\subsection{Junction equation and choice of tension}

The equation of motion of the thin-shell is determined by the junction equation \cite{Israel:1966rt}:
\begin{eqnarray}\label{eq:junc}
\epsilon_{-} \sqrt{\dot{r}^{2}+f_{-}(r)} - \epsilon_{+} \sqrt{\dot{r}^{2}+f_{+}(r)} = 4\pi r \sigma\,,
\end{eqnarray}
where an overdot refers to the derivative with respect to the shell time $t$, $\epsilon_{\pm} = \pm 1$ denote the outward normal direction of outside and inside the shell, respectively, and $\sigma$ is the surface energy density or tension of the shell. It should be noted that the tension $\sigma$ is in general a function of $r$, which satisfies the conservation of the energy that is given by \cite{Blau:1986cw,Ansoldi:1997hz}:
\begin{eqnarray}
\sigma' = - \frac{2(\sigma + \lambda)}{r}\,,
\end{eqnarray}
where $\lambda$ is the surface pressure. The behavior of the tension with respect to $r$ is determined once an equation of state is given (for a study of the thin-shell dynamics in 2+1 dimensions for various equations of state see \cite{Mann:2006yu}). For example, if we assume a constant equation of state, i.e. $w \equiv \lambda/\sigma$, then we obtain
\begin{eqnarray}
\sigma(r) = \frac{\sigma_{0}}{r^{2(1+w)}},
\label{eq:constw}
\end{eqnarray}
where $\sigma_{0}$ is a constant. In addition, if the shell is constructed by a couple of fields with various equations of state, then we can further generalize it as
\begin{eqnarray}
\sigma(r) = \sum_{i=1}^{n} \frac{\sigma_{i}}{r^{2(1+w_{i})}},
\end{eqnarray}
where $\sigma_{i}$s are constants and $w_{i}$ corresponds equations of state.

The most natural realization of this model comes from a scalar field that gives the condition $w=-1$ (and hence a constant tension) \cite{Blau:1986cw}. For the large $r$ limit, this gives the dominant contribution. However, for the small $r$ regime, there could be important contributions from other fields with different equations of states (e.g., \cite{Lee:2015rwa}).

\subsection{Effective potential}

\begin{figure}
\begin{center}
\includegraphics[scale=0.7]{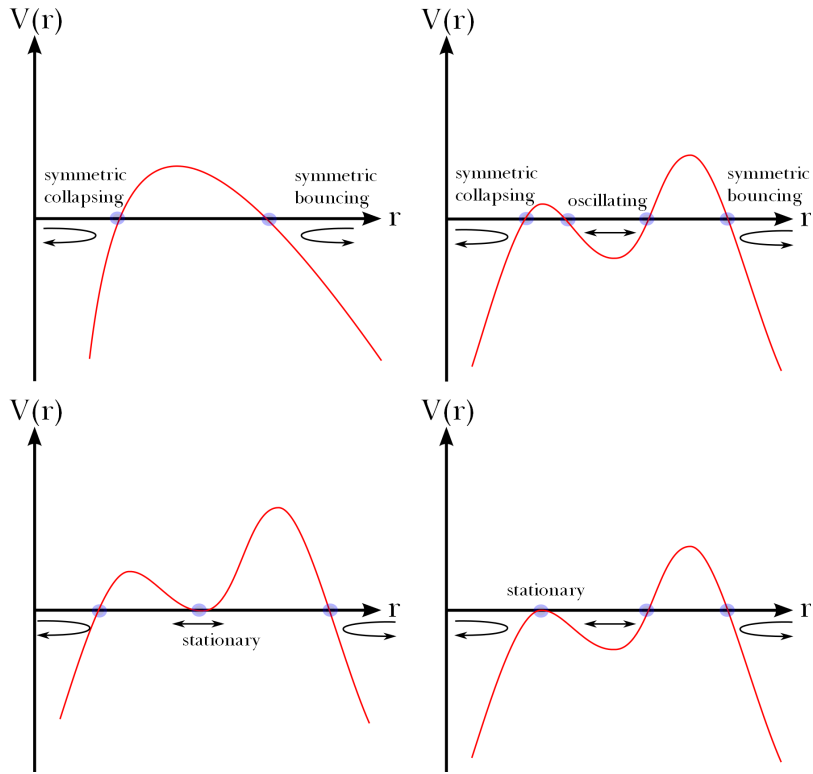}
\caption{\label{fig:pots}Examples of effective potentials, where there are two (upper left), three (lower), and four (upper right) zeros.}
\end{center}
\end{figure}

After some calculations, we reduce the junction equation to a simpler formula \cite{Blau:1986cw}:
\begin{eqnarray}
	\label{eq:form}
\dot{r}^{2} + V(r) &=& 0,\\
V(r) &=& f_{+}(r)- \frac{\left(f_{-}(r)-f_{+}(r)-16\pi^{2} \sigma^{2} r^{2}\right)^{2}}{64 \pi^{2} \sigma^{2} r^{2}}.
\label{eq:form2}
\end{eqnarray}
This effective potential determines the dynamics of the shell, where the classically allowed region is limited by $V(r) \leq 0$. Thus, the number of zeros of the potential especially determines the details. If there is no zero, then the only allowed solution is an \textit{asymmetric} expanding or collapsing shell that connects between $r=0$ and $r=\infty$. However, if there are four zeros (FIG.~\ref{fig:pots}), then these zeros would determine the turning points of the motion of the shell, i.e., where $\dot{r}=0$. Due to presence of these turning points, we find the following possibilities (FIG.~\ref{fig:pots}):
\begin{itemize}
\item[--] \textit{Symmetric collapsing solution}: starting from $r=0$ at $t=-\infty$, approaching the turning point (maximum radius), and finally going to $r=0$ at $t = \infty$.
\item[--] \textit{Symmetric bouncing solution}: starting from $r=\infty$ at $t=-\infty$, approaching the turning point (minimum radius), and finally going to $r=\infty$ at $t = \infty$.
\item[--] \textit{Oscillatory solution}: oscillating between two turning points (a minimum and a maximum radius).
\end{itemize}
In addition, we add two extreme cases of the last possibility where two turning points merge into a single one (lower left and right of FIG.~\ref{fig:pots}):
\begin{itemize}
\item[--] \textit{Stable stationary solution}: a solution where $V(r)=V'(r)=0$ and $V''(r)<0$. Such a stationary solution is classically stable but will eventually either collapse or expand due to quantum effects.
\item[--] \textit{Unstable stationary solution}: if $V(r) = V'(r) = 0$ and $V''(r) > 0$,
 then the solution is stationary at a certain radius, but 
an infinitesimally small perturbation will induce either collapse or expansion;
sometimes such a solution may be relevant as a thermal 
instanton \cite{Gomberoff:2003zh}.
\end{itemize}

In this paper, we mainly focus on a tunneling channel corresponding to a bouncing solution that results in a trivial geometry without a singularity or an event horizon. Previously, several authors have found some examples \cite{Sasaki:2014spa}, but these examples were associated with a symmetric collapsing solution. One potential problem with the symmetric collapsing solution is that it should start from $r=0$ and hence from a singularity and a white hole. It is reasonable to assume that such a region is not physical; if it does not have to satisfy the singularity theorem, then there can be a way to smoothen such an initial condition so that the initial state can be void of singularity.

On the other hand, the initial condition for the gravitational collapse of a real star is very complicated. To avoid such complications, we focus on the role of oscillatory and stationary solutions. These of course are not realistic models of gravitational collapse.
Nevertheless, studying them is still meaningful in that it can explicitly deal with
gravitational collapse geometries from regular initial configurations.

Our task is therefore to find a model where its effective potential $V(r)$ has four zeros to allow for symmetric collapsing solutions, symmetric bouncing solutions, and oscillating solutions; then as extreme cases, we can obtain a stable or unstable stationary solution.

Before going in depth, let us explicitly recover the signs of $\epsilon_{\pm}$ parameters by comparing extrinsic curvatures:
\begin{eqnarray}
	\beta_{\pm}(r) \equiv \frac{f_{-}(r)-f_{+}(r)\mp 16\pi^{2} \sigma^{2} r^{2}}{8 \pi \sigma r} = \epsilon_{\pm} \sqrt{\dot{r}^{2}+f_{\pm}(r)}\,.
	\label{eq:extrinsic}
\end{eqnarray}
It is easy to see that whenever
\begin{align}
	\sqrt{f_--f_+}>4\pi\sigma r\,,
	\label{eq:upper}
\end{align}
$\beta_{+}$ is positive. We will make use of these fact in following sections.

\subsection{Analysis of zeros}

We dedicate this section to analyze the system and to build the previously 
discussed solutions. Essentially, we need to take into account two facts. 
First, the classification 
of solutions is directly related to the number of zeros of the potential. 
Second, the causal structure is determined by the sign of the extrinsic curvature. 
Thus, we are interested in the case when the potential has four zeros 
(or three zeros with one degenerate zero)
and the extrinsic curvatures on both sides of the shell are positive for
 $r$ larger than the event horizon,
 i.e. $\epsilon_\pm=1$ for $r>r_+$ where $f_+(r_+)=0$. 

In other words, we want a stationary or oscillating thin-shell 
to tunnel to a trivial geometry with the resultant shell located at 
somewhere outside of the original shell and expanding to infinity,
which is the one relevant to the future observer. 
Below, we first discuss how to tune the tension
 $\sigma$ so as to fulfill the above conditions. Then we present a couple
of specific examples.

Solutions to $V(r)=0$ under the mentioned requirements are reduced to those satisfying
\begin{eqnarray}
	F(r)\equiv\sqrt{f_{-}} - \sqrt{f_{+}} = 4 \pi \sigma r.
	\label{eq:zeros}
\end{eqnarray}
Thus, if one looks for $n$ zeros, one must choose the tension $\sigma$ 
such that it satisfies Eq.~(\ref{eq:zeros}) $n$ times. 
Let us have a rough intuition on the qualitative behavior of the 
function $F(r)$ in a near horizon, intermediate and far regions. 
We assume $\ell_{\pm}\gg M$ for simplicity.
Near the event horizon $r\gtrsim r_+\sim 2M$, 
since we have $\ell_{\pm} \gg r$, we obtain
\begin{eqnarray}
	F(\ell_{\pm} \gg r) &\simeq&
 1-\sqrt{\frac{r}{2M}-1}+ \mathcal{O}\left(r^{3/2}\right)\,.
\end{eqnarray}
On the other hand, in the intermediate region where $\ell_{\pm}\gg r\gg M$,
a series expansion yields
\begin{eqnarray}
	F(\ell_{\pm}\gg r\gg M) &\simeq& \frac{M}{r}+ \mathcal{O}\left(r^{2}\right)\,.
	\label{eq:behave}
\end{eqnarray}
Lastly, for large $r$, i.e. $r \gg \ell_{\pm}\gg M$, we obtain
\begin{eqnarray}\label{eq:asym22}
	F(r \gg \ell_{\pm}\gg M)&\simeq& \left( \frac{1}{\ell_{-}}
 - \frac{1}{\ell_{+}} \right) r+\mathcal{O}\left(r^{-1}\right)\,,
\end{eqnarray}
where it should be noted that the right-hand side is positive because
$\ell_{-}<\ell_{+}$ for true vacuum bubbles. 
In this way, we have a crude idea on how $F(r)$ behaves and we can 
extract some conditions on the tension.

Before going into details, let us review the asymptotically 
flat case \cite{Sasaki:2014spa}, namely $\ell_+=\infty$, and graphically 
explain the tension building as we base our intuition on this set up. 
First, by a simple analysis, 
it can be shown that in the asymptotically flat case 
one can always find a constant tension which yields two zeros of the potential.
 A \textit{necessary} 
 condition extracted from Eq.~(\ref{eq:asym22}) is
\begin{align}
	\frac{1}{\ell_-}>4\pi\sigma_0\,.
	\label{eq:cond1}
\end{align} 
For the asymptotically AdS case, we expect a similar condition
on $\sigma_0$, namely
\begin{align}
	\frac{1}{\ell_-}-\frac{1}{\ell_+}>4\pi\sigma_0\,.
	\label{eq:cond2}
\end{align} 

\begin{figure}
	\begin{center}
		\includegraphics[scale=0.7]{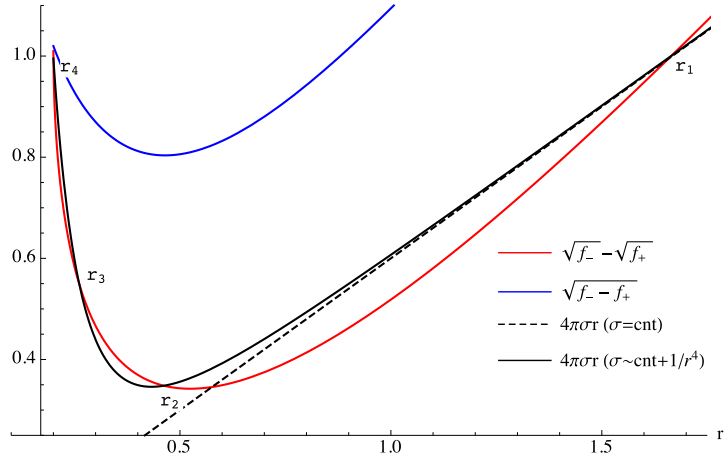}
		\caption{\label{fig:qualitative} Qualitative behavior of $\sqrt{f_{-}} - \sqrt{f_{+}}$ (red), 	$\sqrt{f_{-} - f_{+}}$ (blue) and $4\pi\sigma r$ (black) for true vacuum bubbles in AdS background. A constant tension $\sigma$ (black dashed line) can give rise to two zeros of the potential at most. Four zeros could be obtained by modifying the tension at small $r$ (black line). The blue line sets the upper bound for the tension so as not to change the sign of the extrinsic curvature.}
	\end{center}
\end{figure}

As seen explicitly from FIG.~\ref{fig:qualitative}, 
a constant tension line (dashed black line) crosses $F(r)$ (red curve) twice,
one in the large $r$ region and the other in the $\ell_{\pm}\gg r\gg M$ region
where $F(r)$ starts to behave as $1/r$. Let us denote the positions
of these two zeros by $r_1$ and $r_2$, where $r_1>r_2$.
 The blue curve shows the upper bound on $4\pi\sigma r$ so that the 
extrinsic curvature is positive at any $r$ greater than $r_{+}$
(see Eq.~(\ref{eq:upper})). That ensures the presence of two zeros. 
In fact, since $\sqrt{f_{-}-f_{+}}\geq F(r)$ for $r\geq r_+$ and
the equality holds only at the horizon, the existence of a zero
in the large $r$ region guarantees that of another zero in the intermediate region.

Thus we first choose the constant part of the tension that satisfies 
Eq.~(\ref{eq:cond2}). Afterwards, we modify the tension at $r<r_2$, 
around the $1/r$ region, so as to have the resulting curve (black curve)
intersect the red curve two more times, as shown in FIG.~\ref{fig:qualitative}. 
This can be achieved if one choose the tension to behave as $\sigma\sim1/r^\alpha$ 
with $\alpha>2$ at $r<r_2$ and, hence, the product $\sigma r$ has
a steeper gradient than $F(r)$ as $r$ decreases.
This condition is equivalent to 
requiring the equation of state of the shell to cross the 
value $w=-1/2$, as one can see from evaluating
\begin{eqnarray}\label{eq:sigmaprime}
	(\sigma r)'=-\sigma(1+2w).
\end{eqnarray}

\begin{figure}[h]
	\begin{center}
		\includegraphics[scale=0.5]{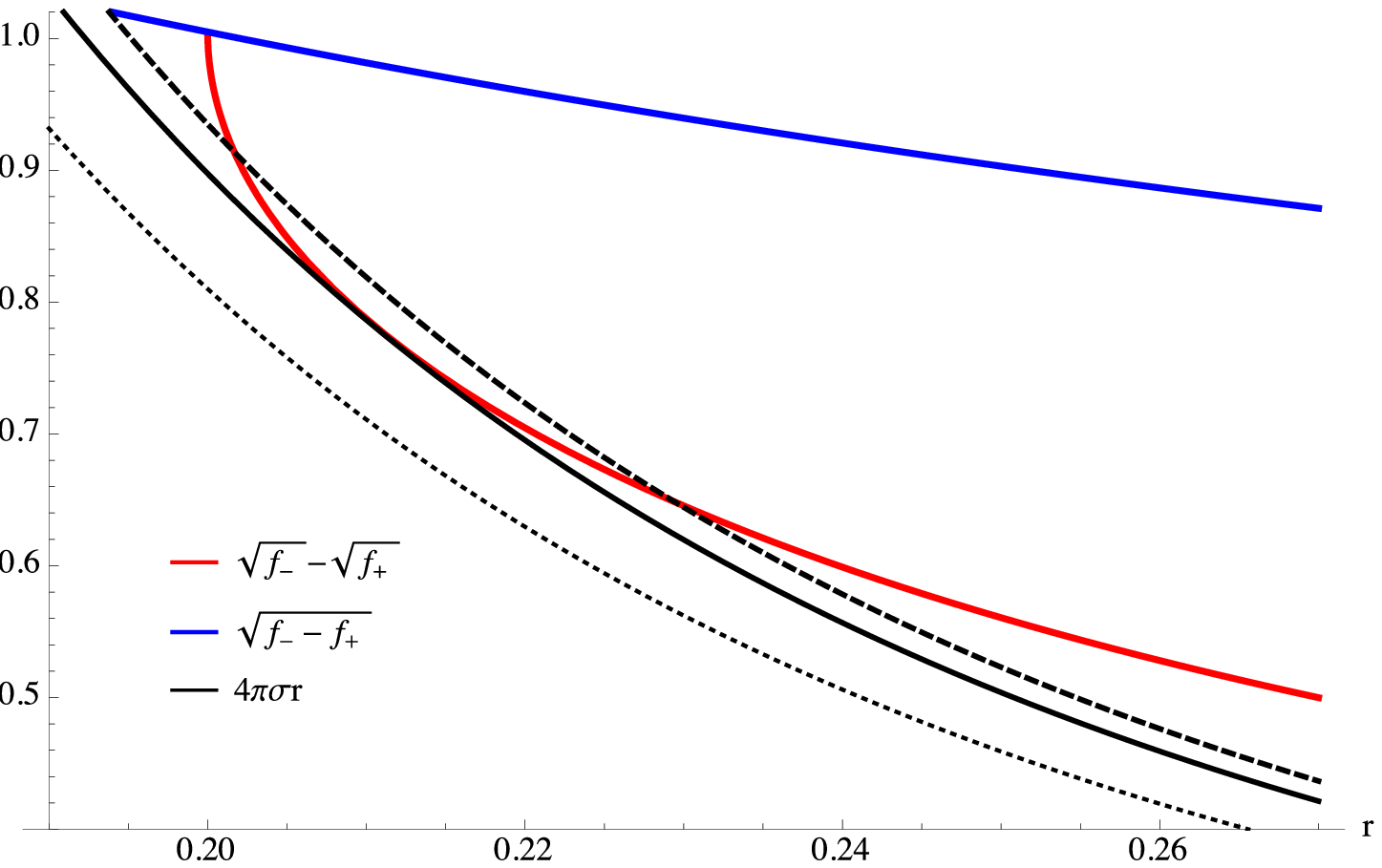}
		\includegraphics[scale=0.5]{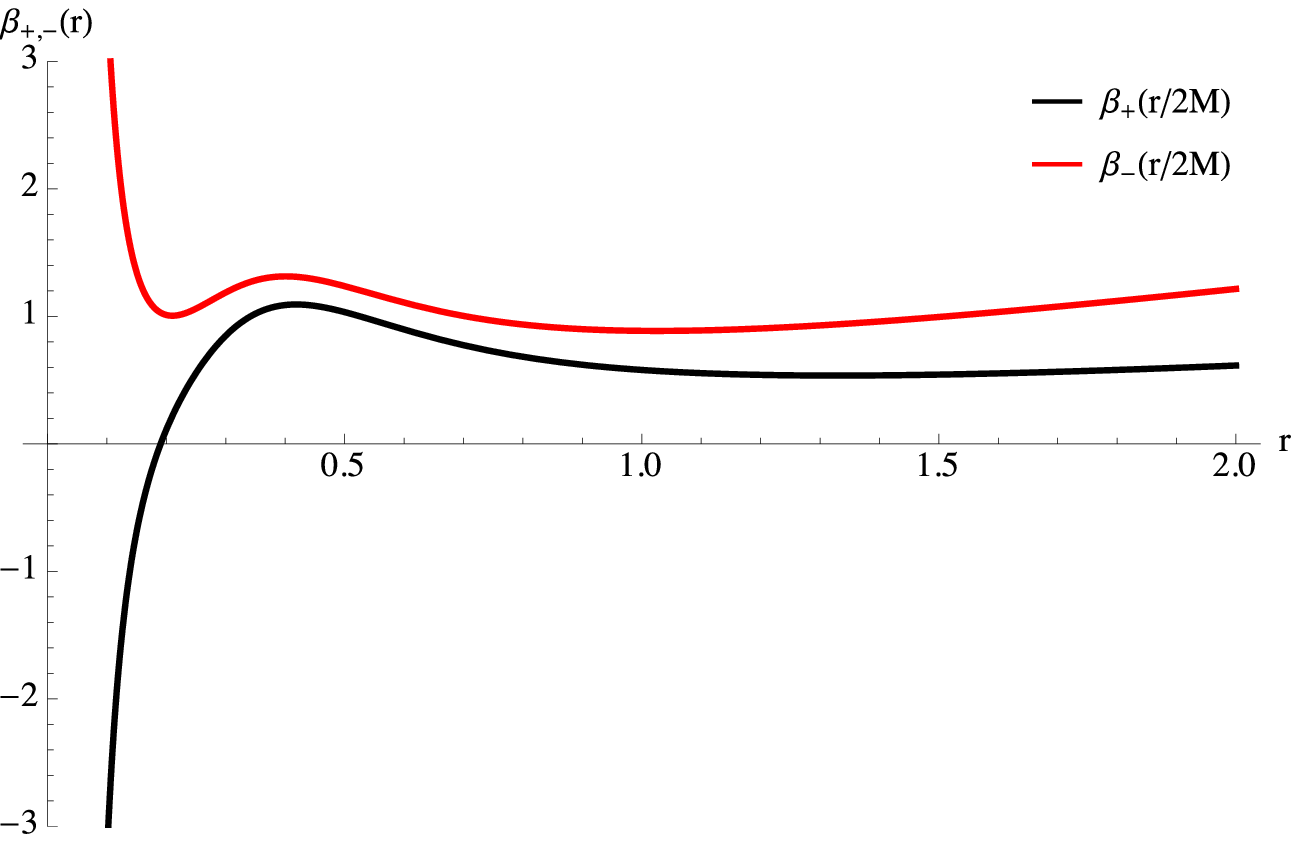}
		\caption{\label{fig:extcurv} Left. $\sqrt{f_{-}}-\sqrt{f_{+}}$ (red) and $\sqrt{f_{-}-f_{+}}$ (blue) when $M=0.1$, $\ell_{-}=2$, $\ell_{+}=\infty$ and $4\pi\sigma_{0} = 0.3$. $4\pi\sigma_{1}$ runs from $0.006$ (bottom, dotted), $0.0067$ (curve), $0.007$ (dashed), to $0.008$ (top, dotted). 
			Right. The corresponding extrinsic curvatures $\beta_{+}$ (black) and $\beta_{-}$ (red) for $4\pi\sigma_{1} = 0.0067$.}
		\includegraphics[scale=0.8]{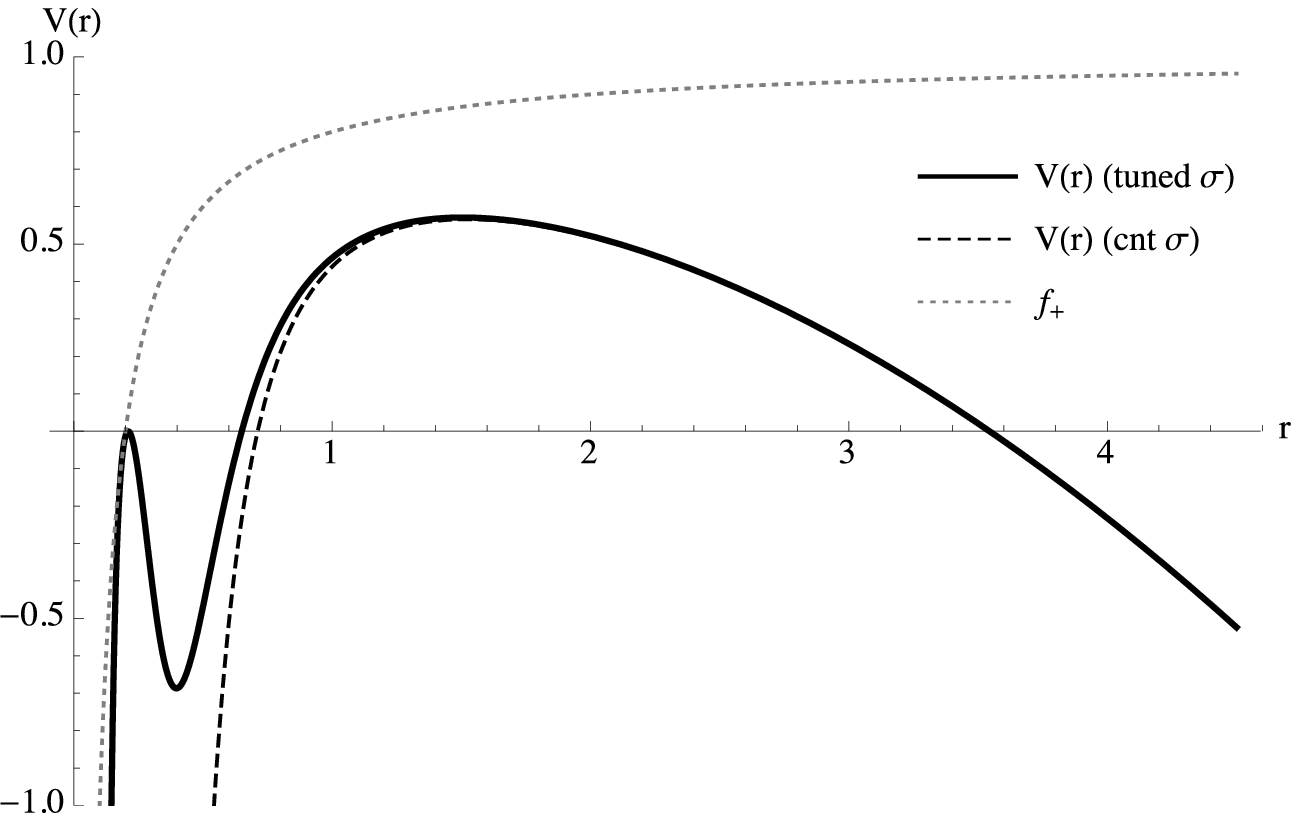}
		\caption{\label{fig:effpot}$V(r)$ for $M=0.1$, $\ell_{-}=2$, $\ell_{+}=\infty$, $4\pi\sigma_{0} = 0.3$ and $4\pi\sigma_{1} = 0.0067$. The dashed curve indicates the case when $\sigma_{1}=0$ and $f_+(r)$ (dotted line) is shown as a cross check that both extrinsic curvatures are positive for $r>r_+$.}
	\end{center}
\end{figure}

\paragraph{A concrete example: an unstable stationary thin-shell.}
One can easily find an example that has an unstable stationary solution. 
Let us consider a tension given by
\begin{eqnarray}
	\sigma(r) = \sigma_{0} + \frac{\sigma_{1}}{r^{4}}\,,
\end{eqnarray}
where  $\sigma_{0}$ ($w=-1$) is set from Eq.~(\ref{eq:cond2}) and that the potential has two zeros as explained above. Therefore, we have to choose the contribution $\sigma_1$ from stiff matter ($w=1$) such that two more zeros at the small $r$ regime arise.
Let us call the latter \textit{Model 1}.
 FIGs.~\ref{fig:extcurv} and \ref{fig:effpot} show a particular example 
when $M=0.1$, $\ell_{-}=2$, $\ell_{+}=\infty$ and $4\pi\sigma_{0} = 0.3$. 

Concretely, the left panel of FIG.~\ref{fig:extcurv} shows $F(r)$ (red) and 
the upper bound on $4\pi\sigma r$ (blue) at small $r$. The two curves meet 
at the apparent horizon $r=0.2$. Here, we superimpose with black curves 
different values of $\sigma_1$, specifically $4\pi\sigma_{1}=0.006$
 (bottom, dotted), $0.0067$ (curve) and $0.007$ (dashed).

In the critical limit where $4\pi\sigma_{1} = 0.0067$, the potential has 
three zeros and hence it has a unstable stationary point. This can be 
clearly checked in terms of the effective potential (see FIG.~\ref{fig:effpot}).
Additionally, the right panel of FIG.~\ref{fig:extcurv} confirms that the extrinsic
 curvatures are always positive for $r>r_+$.

\paragraph{A special case: thin-shell in pure vacuum background.}

Let us turn our attention into the particular case when the vacuum
energy in both inside and outside the shell vanishes, i.e. $\ell_\pm=\infty$.
We are interested in building a classical stable thin-shell which may either
tunnel to a black hole or expand to infinity leaving flat spacetime behind
without any singularity. However, the foregoing discussion on true
vacuum bubbles in asymptotically AdS background does not apply any longer
as the function $F(r)$ approaches zero as $1/r$ in the limit $r\to\infty$
(see Eq.~(\ref{eq:behave})).
Thus, we need to tune the tension in a different way.

As $F(r)\propto 1/r$ for $r\to\infty$, the tension should
decrease faster than $1/r^2$ to realize a zero at large $r$
(ie, for $4\pi\sigma r$ to cross $F$ from above).
This requires $w>0$ as seen from Eq.~(\ref{eq:constw}). 
On the other hand, since $F\sim 1-\sqrt{(r-2M)/2M}$ near the horizon, 
any monotonically decreasing $\sigma$ which satisfies $8\pi M\sigma<1$ 
and $d(\sigma r)/dr<0$ at $r=2M$ would have a zero near the horizon
(i.e., $4\pi\sigma r$ crosses $F$ from below), which requires $w>-1/2$. 

The above consideration implies that there would be two zeros for
a constant $w$ with $w>0$. However, we need two more zeros. This can
be realized if we set $w$ sufficiently large near the horizon so that
$\sigma r$ decreases faster than $F$ in a region slightly away from
the horizon. Below we propose a model that satisfies these conditions.

The effective potential in the present case is given by
\begin{eqnarray}
	V(\bar{r})=1-\frac{1}{\bar{r}}
-\frac{1}{4}\left({\frac{1}{\bar{\sigma}\bar{r}^{2}}}-\bar{\sigma}\bar{r} \right)^{2}\,,
\end{eqnarray}
where we normalized $r$ to the Schwarzschild radius, i.e. $\bar{r}=r/2M$, 
and defined $\bar{\sigma}\equiv8\pi M\sigma$. In this way, the analysis 
is effectively independent of $M$. 
In this case, the requirement that the extrinsic curvatures are positive translates to
\begin{eqnarray}\label{eq:condition}
	\bar{\sigma}^{2} \bar{r}^{3}<1.
\end{eqnarray}
This is fulfilled near the horizon, $\bar{r}\sim1$, if $\bar{\sigma}<1$ 
and also in the large $r$ limit if $w > 0$. 
Among many possibilities of tensions which fulfill these criteria, 
 we consider the one which seems to be physically motivated as 
well as simple.

The model is as follows.  When the thin-shell tunnels and expands 
to infinity, only radiation will reach the observer. So we may assume 
 $w(\bar{r}\rightarrow\infty)=1/3$. On the other hand, 
as discussed in the above, the tension $\sigma$ must increase fast enough 
to stabilize the shell as $r$ decreases. Thus, we consider a shell that
undergoes a phase transition to stiff matter as $r$ decreases, 
that is $w(\bar{r}\rightarrow0)=1$. 
Furthermore, we assume the thin-shell is composed of dust matter 
in the intermediate region, i.e. $w=0$.

To realize the above behavior, let us propose
\begin{align}
	w(\bar r)=\left\{
	\begin{array}{ccc}
		\frac{1}{2}\tanh\left[\mu_1\left(\bar{r}_1-\bar{r}\right)\right] +\frac{1}{2}& \quad \bar{r}<\bar{r}_m
		\\
		\frac{1}{6}\tanh\left[\mu_2\left(\bar{r}-\bar{r}_2\right)\right] +\frac{1}{6} & \quad \bar{r}>\bar{r}_m
	\end{array}
	\right.,
\end{align}
where $\mu_i$ and $r_i$ control, respectively, the steepness and position of 
the transition, and $r_m$ is the matching radius in the dust region. 
Notice that if the two transitions are sufficiently far apart,
i.e. $\bar r_1\ll \bar r_m \ll \bar r_2$, the matching is automatic
and the exact value of the matching radius $r_m$ becomes irrelevant.
Lastly, we set the tension as
\begin{align}
	\bar\sigma(\bar r)=\bar{\sigma}_0\,\bar{r}^{-(2+2w(\bar{r}))}\,
\end{align}
where $\bar\sigma_0<1$. We note that because of the $\bar{r}$-dependence
of $w$, it is no longer exactly equal to the ratio of the pressure to the energy
density. Nevertheless, for sufficiently steep transitions $w$ is a
good approximation to the pressure to the energy density ratio almost everywhere. 

\begin{figure}
	\begin{center}
		\includegraphics[scale=0.5]{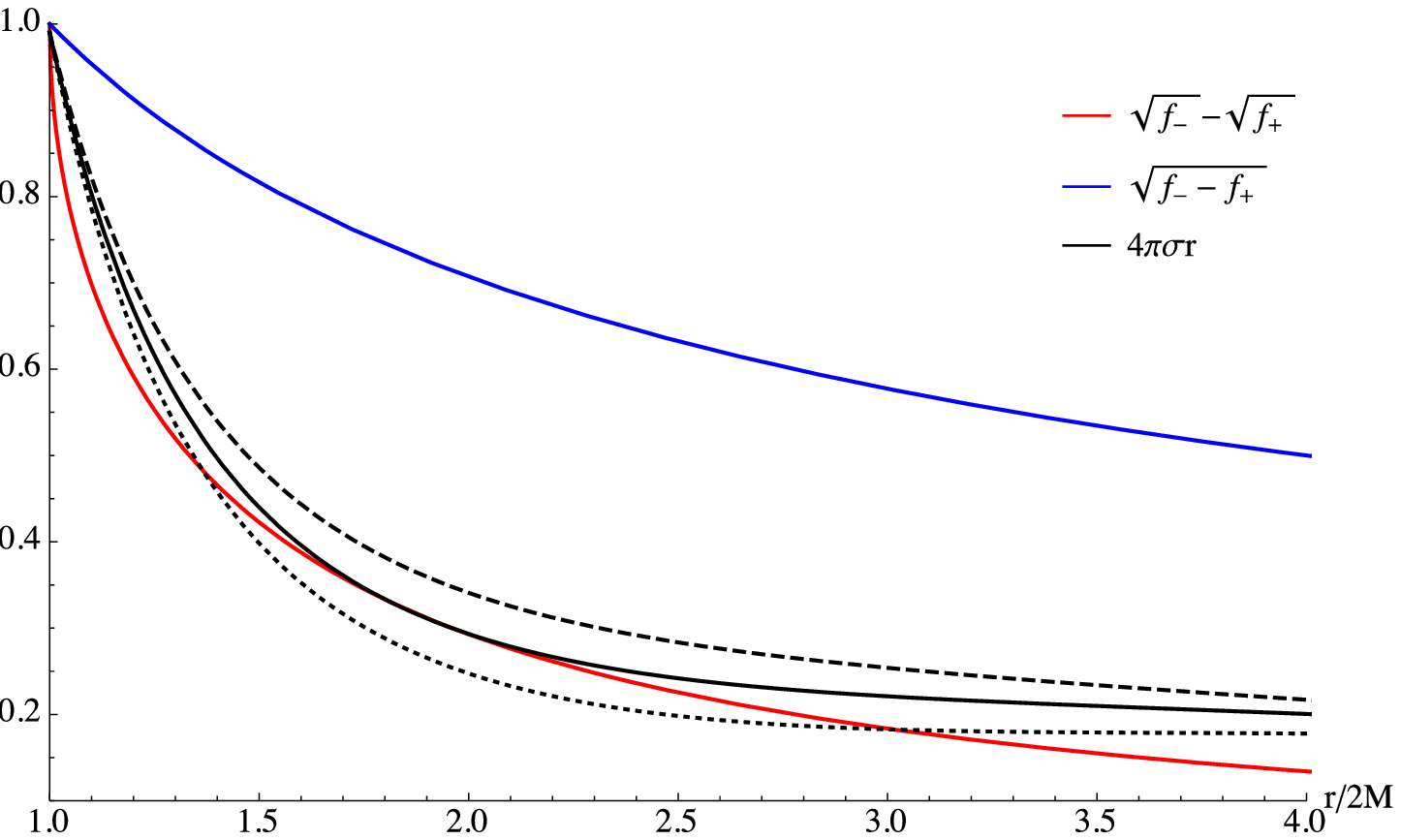}
		\includegraphics[scale=0.5]{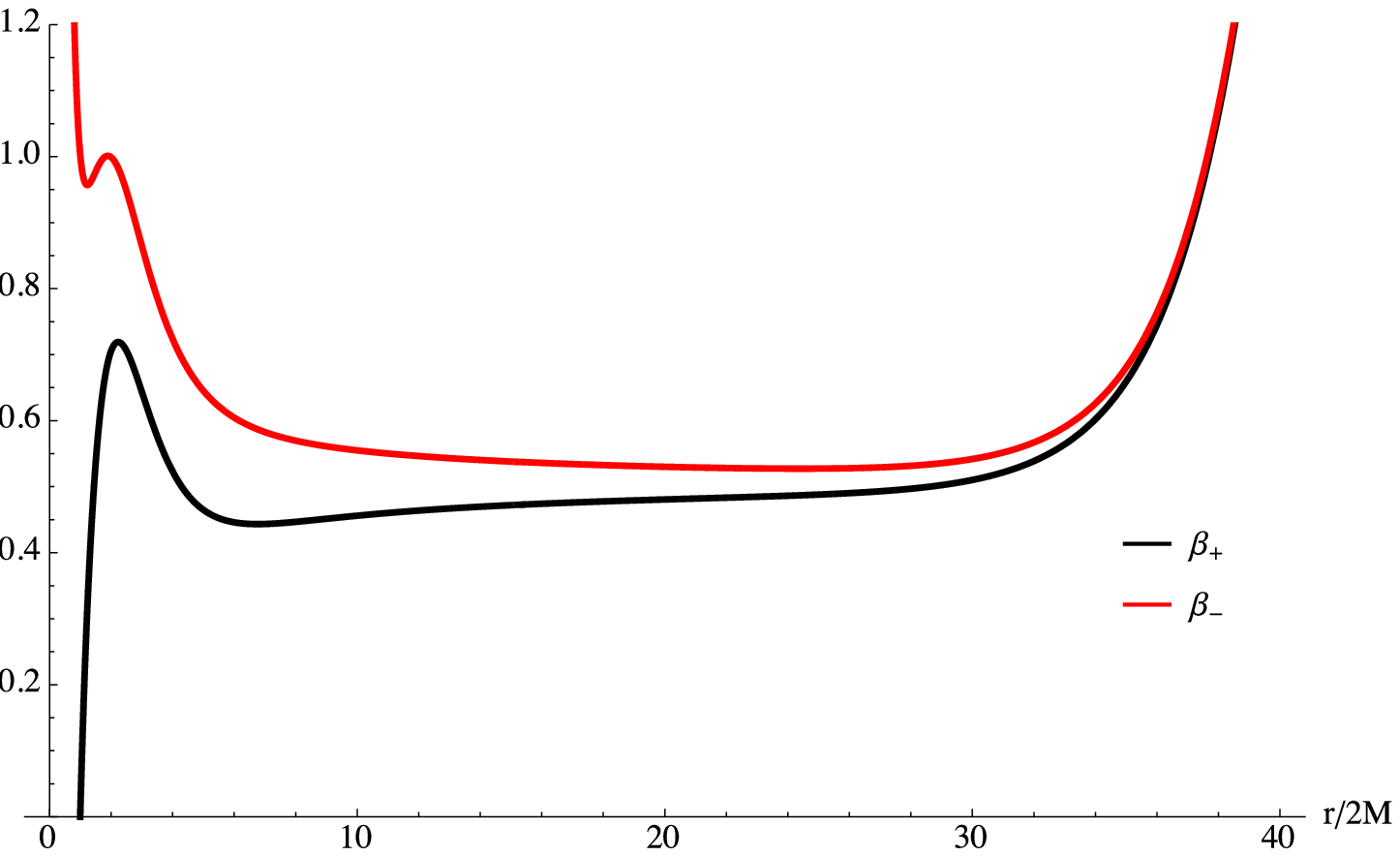}
		\caption{\label{fig:sch1}Left. $\sqrt{f_{-}}-\sqrt{f_{+}}$ (red), $4\pi\sigma r$ (black) of Model~2 when $\bar\sigma_0=0.99$, $\mu_{1}=1/2$, $\bar{r}_{1}=1.5$, $\mu_{2}=1/5$ and $\bar{r}_{2}=40$. The additional black lines are $\bar{r}_{1}=2$ (dotted) and $\bar{r}_{1}=1$ (dashed). It should be noted that by construction $4\pi\sigma r$ crosses again the curve $\sqrt{f_{-}}-\sqrt{f_{+}}$ and, therefore, the total number of zeros is three. Right. Extrinsic curvatures of Model~2, black for $\beta_{-}$ and red for $\beta_{+}$. Both of them are positive for $r>r_+$.}
		\includegraphics[scale=0.8]{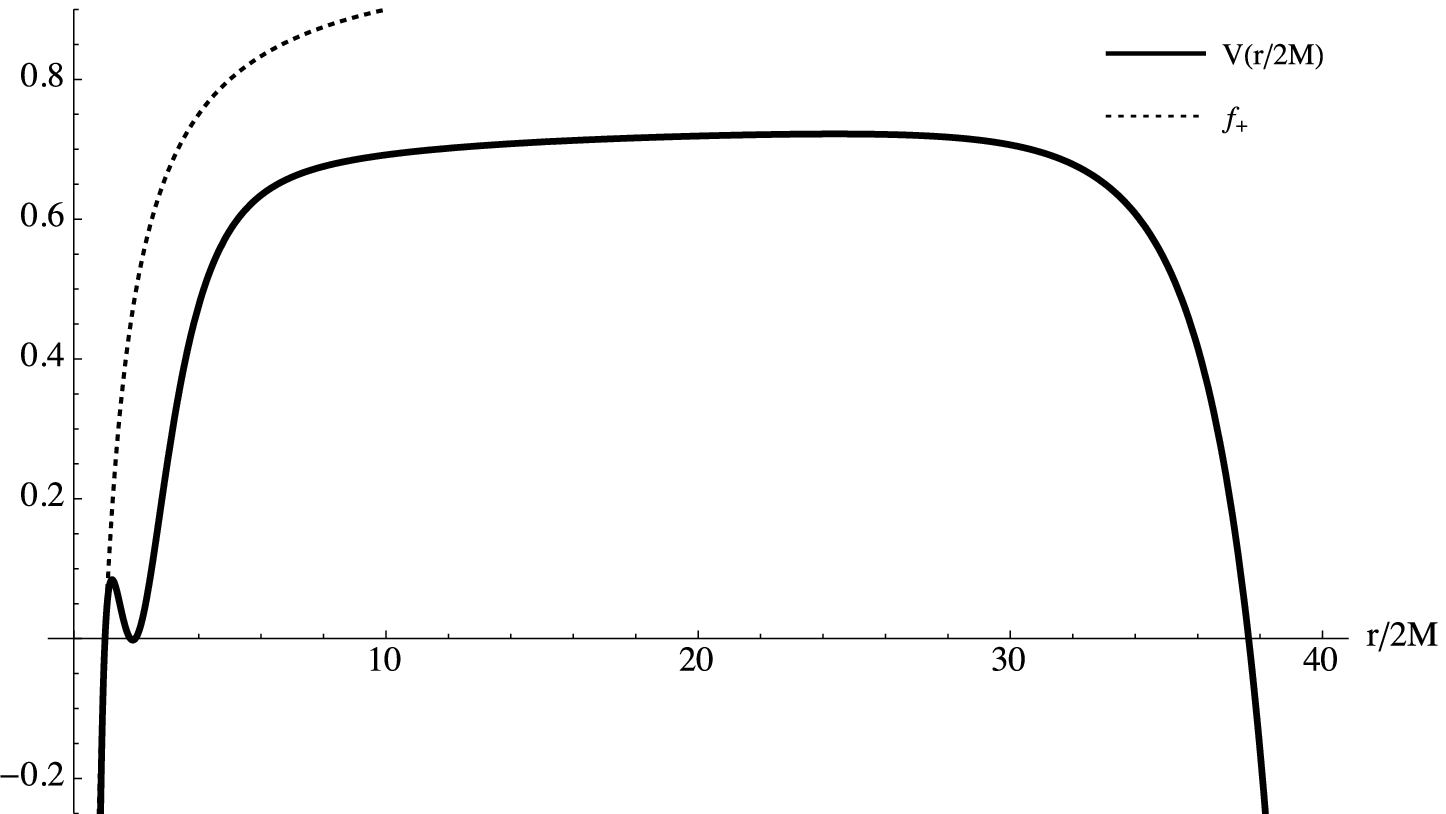}
		\caption{\label{fig:sch2}This figure shows $V(r)$ and $f_+(r)$. The shell is located in a local stable minimum and can either collapse or expand by tunneling effects. Notice that there is no point for $r>r_+$ where $V(r)=f_+(r)$ and, hence, the extrinsic curvatures are both positive.} 
	\end{center}
\end{figure}

As a concrete model, we first choose $\bar\sigma_0<1$ so that the positivity
of the extrinsic curvatures is guaranteed. Next we require that the 
transitions are fast enough, which roughly implies 
$\mu_1\gtrsim \mathcal{O}(1)$ and $\mu_2\gtrsim \mathcal{O}(1)$.
Afterwards, we choose $\bar r_1$ and $\bar r_2$. Here we consider
a set of models with various values of $\bar r_1$ which are
not too large compared to unity and with a fixed value of $\bar r_2$
which is large enough so that the matching is smooth enough,
 i.e. $|\mu_1(r_1-r_m)|\gg1$ and $|\mu_2(r_2-r_m)|\gg1$.
A particular example is shown in FIGs.~\ref{fig:sch1} and \ref{fig:sch2} 
where we choose $\bar\sigma_0=0.99$, $\mu_{1}=1/2$, $\mu_{2}=1/5$ 
and $\bar{r}_{2}=40$. We refer to it as \textit{Model 2}. 
If $\bar{r}_{1}=1.5$ then one has three zeros and, therefore, 
a stable stationary thin-shell.

\begin{figure}
	\begin{center}
		\includegraphics[scale=0.75]{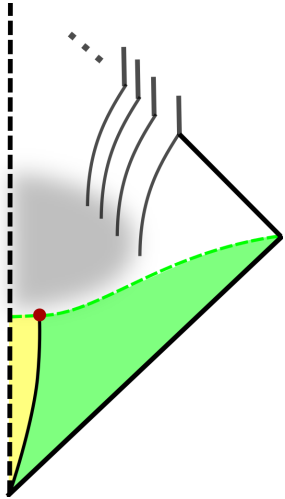}
		\caption{\label{fig:tunnelinggeos2}After summing over histories, we see the superposed geometry $\langle g_{\mu\nu} \rangle$.}
	\end{center}
\end{figure}

\subsection{Tunneling probability}
Before ending this section, let us compute the tunneling probability 
for Model~2. As usual, in the Euclidean path integral approach, 
the probability is estimated semi-classically by the WKB approximation, that is
	\begin{align}
		P\sim{\rm e}^{-B/\hbar}\,,
	\end{align}
	where $\hbar$ is the reduced Planck constant and 
	\begin{align}
		B=2\int_{r_{\times a}}^{r_{\times b}} |P_{\mathrm{E}}(r)|dr
	\end{align}
	is the integral of the Euclidean momentum of the shell between two consecutive zeros, $r_{\times a}$ and $r_{\times b}$, which is given by \cite{Ansoldi:1997hz}
	\begin{align}
		P_{\mathrm{E}}(r)=-r\cos^{-1}\left(\frac{f_-+f_+-(4\pi\sigma r)^2}{2\sqrt{f_-f_+}}\right)\,,
	\end{align}
	where $-\pi<\cos^{-1}<0$. In fact, due to the simplicity of our example
 in the pure vacuum background, we find that the probability is given by
	\begin{align}
		\frac{B}{\hbar}=\frac{M^2}{M_{\mathrm{Pl}}^2}\frac{1}{\pi}\int_{\bar{r}_{\times a}}^{\bar{r}_{\times b}} |P_{\mathrm{E}}(\bar{r})|d\bar{r}\,,
	\end{align}
where we have recovered the units and $M_{\mathrm{Pl}}$ is the reduced 
Planck mass ($=\sqrt{\hbar c/8\pi G}$). It should be noted that the integral 
is just a numerical value and that, for our choice of tension,
 a rough numerical approximation to the integral for the tunneling
 to a black hole and an expanding shell respectively gives
	\begin{align}
		& B_{\rm BH}\approx0.3~{M^2}/{M_{\mathrm{Pl}}^2}\qquad {\rm and}\qquad
		B_{\rm expand}\approx27.4~{M^2}/{M_{\mathrm{Pl}}^2}\, ,
	\end{align}
respectively. Note that the former is $100$ times small than the latter. 
This result agrees with the well believed fact that the tunneling 
probability is exponentially suppressed with the exponent of the order
of the entropy. More explicitly, 
the ratio between probabilities for our case yields
	\begin{align}
		\frac{P_{\rm expand}}{P_{\rm BH}} \sim {\rm e}^{-27.1{M^2}/{M_{\mathrm{Pl}}^2}}\,,
	\end{align}
	namely that the probability that the shell tunnels into a black hole is exponentially higher.
	Regarding the information loss paradox, if one thinks of this stable thin-shell as a quantum mechanical system, no actual issue arises. Basically, one has a wave packet which is located at the stable minimum of the potential and which will evolve and spread. Of course, the highest probability goes towards the formation of a black hole but there is spreading of the wave function for large $r$ as well. In this sense, there is no information loss.
	
	Interestingly, in the limiting case where $M\sim \mathcal{O}(M_{\mathrm{Pl}})$, if we can give any physical meaning to it, the probability of an expanding solution is no longer exponentially suppressed and the superposition of geometries would start to be important, illustratively shown in FIG.~\ref{fig:tunnelinggeos2}. In any case, we have to bear in mind that this is a particular example and the conclusion depends on the choice of the tension. Nevertheless, this particular example could be considered as
a model for gravitational stellar evolution where the most probable 
final stage is to form a black hole.

\section{\label{sec:con}Conclusion}

In this paper, we investigate stationary bubbles and their tunneling channels 
 a trivial geometry. By tuning the tension of the shell under reasonable 
assumptions on the equation of state, we realize stationary solutions.
 These stable and unstable stationary shells could be a good toy model 
 for a star interior around the time when gravitational collapse turns on. 
In this way, it can either collapse to a black hole or tunnel toward an expanding 
 shell. For the latter case, even though the probability is exponentially
 suppressed, there will be no singularity nor event horizon, and hence
 information is conserved under the Euclidean path integral approach
 to the wave function of the universe.

What we achieved in this paper can be summarized as follows:
\begin{itemize}
\item[--] In the time-symmetric solution, there is a part of the 
initial singularity and the white hole horizon. However, 
the initial singularity is \textit{not} the essential feature; 
we can explicitly construct a model that does not start from an 
initial singularity.
\item[--] For realistic gravitational collapse, there can be much 
 diverse (maybe, infinite number of) branching channels toward a 
trivial geometry, where each process will be exponentially suppressed
 with the exponent given by the order of the entropy $\sim M^{2}/M_{\rm Pl}^2$.
\item[--] We constructed a classically stable thin-shell example. 
In this case, the probability that forms a black hole and that expands
 to infinity will be \textit{both} exponentially suppressed
 (though that depends on the tension). The two probabilities will be 
unsuppressed and of the same order when the mass becomes the Planck scale, 
i.e., when the ``fuzziness'' of the geometry becomes important.
\end{itemize}

We found various parameters that allow both collapsing and bouncing 
solutions. While this may not be the most general case, these 
parameters seem quite natural and therefore allowed. This suggests 
that we may go one step forward toward a \textit{generic} behavior of
 the wave function of the universe. One must bear in mind that we are
 dealing with a broad class of examples but still it is no definite
 proof that there always exists an instanton that mediates a trivial
geometry. However, if this generic statement were to be proven someday, 
then we might reach a solution to the information loss paradox through
 the Euclidean path integral approach.

\section*{Acknowledgment}
GD and MS would like to thank S. Ansoldi for many insightful comments. PC is supported by National Center for Theoretical Sciences (NCTS) and Ministry of Science and Technology (MOST) of Taiwan. GD and MS are supported by MEXT KAKENHI Grant Number 15H05888. DY is supported by Leung Center for Cosmology and Particle Astrophysics (LeCosPA) of National Taiwan University (103R4000). The authors would like to thank hospitality from Molecule-type Workshop on Black Hole Information Loss Paradox (YITP-T-15-01), held in Yukawa Institute for Theoretical Physics, Kyoto University, Japan.

\newpage

\section*{Appendix: Causal structure and interpretation}

\begin{figure}
	\begin{center}
		\includegraphics[scale=0.5]{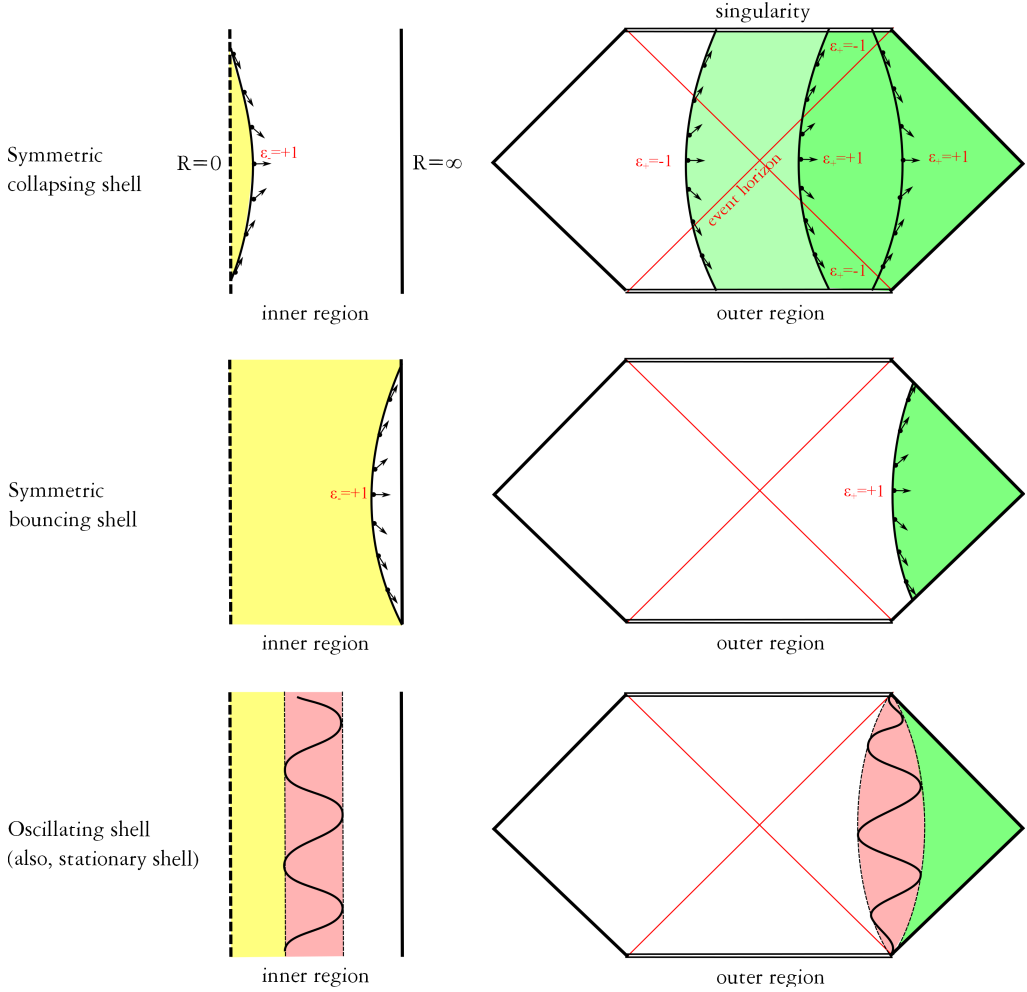}
		\caption{\label{fig:signs_Mink}Causal structures of symmetric collapsing solutions (upper), symmetric bouncing solutions (middle), and oscillating solution (lower), where left is the inner AdS and right is the outer Schwarzschild (we can easily generalize to asymptotic AdS or dS).}
	\end{center}
\end{figure}

In this appendix we give an illustrative interpretation of the causal structures of Model 1 and Model 2, see FIG.~\ref{fig:signs_Mink}. For symmetric collapsing solutions, in Model~1, there are two kinds: either there exists a region where $\epsilon_{+} = +1$ or $\epsilon_{+}$ is always negative; in Model~2, always $\epsilon_{+} = +1$ and the trajectory should follow the condition. If the bubble is a true vacuum bubble and if there is a symmetric bouncing solution, then the only possible solution is $\epsilon_{\pm} = +1$. Finally, we are interested in an oscillatory solution that is located the right side of the causal patch. A stable stationary solution is a special case when the turning points are close each other.

\begin{figure}
	\begin{center}
		\includegraphics[scale=0.75]{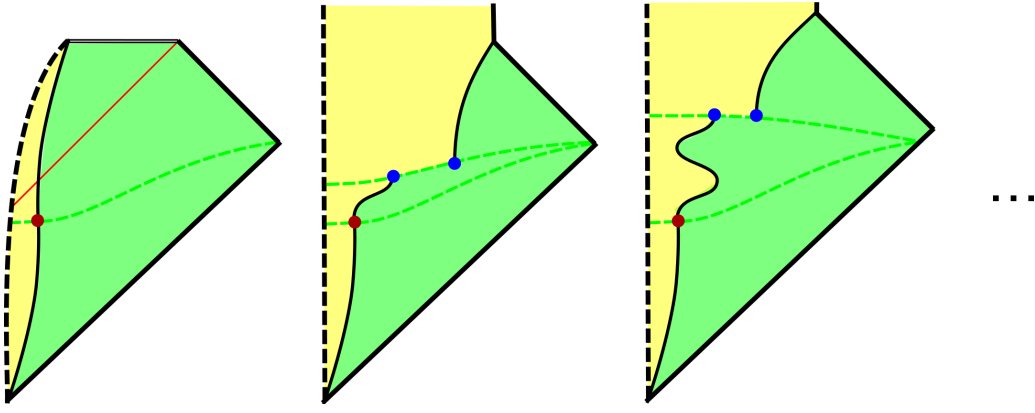}
		\caption{\label{fig:tunnelinggeos}\textit{Left}: an initial unstable shell begins to collapse and form a black hole (for Model~1). For Model~2, the shell will bend toward the left direction around the singularity. \textit{Middle}: an initial unstable shell begins to bounce and tunnels toward a trivial geometry. \textit{Right}: there can be various tunneling channels.}
	\end{center}
\end{figure}

We can construct possible tunneling channels (FIG.~\ref{fig:tunnelinggeos}) so-called Farhi-Guth-Guven/Fischler-Morgan-Polchinski tunneling \cite{Farhi:1989yr}. This figure assumed the case when initially the shell was in the unstable local maximum. If the shell was in the stable local minimum, to trigger a formation of a black hole, one requires a tunneling and this will be easily generalized from FIG.~\ref{fig:tunnelinggeos}. The unitary observer should superpose every histories (FIG.~\ref{fig:tunnelinggeos2}) \cite{Hartle:2015bna}; in terms of such an observer, the averaged metric $\langle g_{\mu\nu} \rangle$ can look like fuzzy that may not satisfy classical equations of motion (hence general relativity).

\end{document}